\documentclass{article}

\usepackage{arxiv}

\usepackage[utf8]{inputenc} % allow utf-8 input
\usepackage[T1]{fontenc}    % use 8-bit T1 fonts
\usepackage{hyperref}       % hyperlinks
\usepackage{url}            % simple URL typesetting
\usepackage{booktabs}       % professional-quality tables
\usepackage{nicefrac}       % compact symbols for 1/2, etc.
\usepackage{microtype}      % microtypography
\usepackage{lipsum}
\usepackage{amsmath,amssymb,amsfonts}
\usepackage{algorithmic}
\usepackage{graphicx}
\usepackage{textcomp}
\usepackage{xcolor}
\usepackage{hyperref}

\title{Heterogeneous Reservoir Computing Models for Persian Speech Recognition}

\author{
 Zohreh Ansari \\
  Biomedical Engineering Department\\
  Meybod University\\
  Meybod, Iran \\
  \texttt{z\_ansari@meybod.ac.ir} \\
   \And
 Farzin Pourhoseini \\
  Biomedical Engineering Department\\
 Meybod University\\
 Meybod, Iran\\
  \texttt{pourhoseinifarzin@gmail.com} \\
  \And
 Fatemeh Hadaeghi \\
  Institute of Computational Neuroscience\\
  University Medical Center Hamburg-Eppendorf (UKE)\\
  Hamburg, Germany\\
  \texttt{f.hadaeghi@uke.de} \\
}

\begin{document}
\maketitle
\begin{abstract}
Over the last decade, deep-learning methods have been gradually incorporated into conventional automatic speech recognition (ASR) frameworks to create acoustic, pronunciation, and language models. Although it led to significant improvements in ASRs' recognition accuracy, due to their hard constraints related to hardware requirements (e.g., computing power and memory usage), it is unclear if such approaches are the most computationally- and energy-efficient options for embedded ASR applications. Reservoir computing (RC) models (e.g., echo state networks (ESNs) and liquid state machines (LSMs)), on the other hand, have been proven inexpensive to train, have vastly fewer parameters, and are compatible with emergent hardware technologies. However, their performance in speech processing tasks is relatively inferior to that of the deep-learning-based models. To enhance the accuracy of the RC in ASR applications, we propose heterogeneous single and multi-layer ESNs to create non-linear transformations of the inputs that capture temporal context at different scales. To test our models, we performed a speech recognition task on the Farsdat Persian dataset. Since, to the best of our knowledge, standard RC has not yet been employed to conduct any Persian ASR tasks, we also trained conventional single-layer and deep ESNs to provide baselines for comparison. Besides, we compared the RC performance with a standard long-short-term memory (LSTM) model. Heterogeneous RC models (1) show improved performance to the standard RC models; (2) perform on par in terms of recognition accuracy with the LSTM, and (3) reduce the training time considerably.
\end{abstract}

% keywords can be removed

\section{Introduction}
\subsection{Practical relevance and motivation}
\label{sec:motivation}

Over decades of research, hidden Markov models and different variations of neural networks have been extensively exploited to perform speech recognition, speaker identification, text to speech conversion, and other tasks that are relevant for speech processing applications \cite{ganapathiraju2004applications,shannon2003comparative,gales2008application,morgan2011deep}. During the past decade, and thanks to remarkable advancements in graphics processing unit (GPU) and cloud computing technologies, a large variety of deep-learning-based methods have been designed and tested to accomplish challenging speech processing tasks on large datasets \cite{dahl2011context,de2018neural,hinton2012deep,yu2012deep}. In conventional ASR systems, these models are primarily employed to extract features, to uncover the relationships between the audio signal and the phonemes (or other linguistic units), to learn pronunciation lexicons, and to recognize (and use) sequences of words. These developments, however, have been mainly focused on improving recognition accuracy by training models on a multitude of collected data. Imprecision and false interpretation, although momentous, are not the only challenges faced by practical ASR systems. Since algorithms have to be computationally efficient, less data-hungry, and compatible with emerging micro-device technologies, further investigations are needed to develop inexpensive yet accurate processing methods. Besides, it is of crucial importance to make speech recognition available for more languages and dialects. 

In contrast to deep-learning-based methods, reservoir computing (RC) models (e.g., echo state networks (ESNs) \cite{jaeger2004harnessing} and liquid state machines (LSMs) \cite{maass2002real}) have been proven inexpensive to train, have vastly fewer parameters, and reported to perform well in processing complex temporal and spatio-temporal data in real-world applications \cite{hadaeghi2021spatio,triefenbach2010phoneme,lukovsevivcius2012reservoir}. It has also been suggested that RC models can provide accurate subject-specific classifiers that are adaptable to the unique characteristic features of temporal data recorded from a target person and do not rely on a vast amount of data collected from other individuals \cite{hadaeghi2019reservoir}. Besides, the essential ingredient of a reservoir computing model is a random excitable medium that non-linearly projects an input signal into a higher-dimensional signal space. Therefore, researchers from computing theory and microchip technologies have considered RC as a computational scheme compatible with ``unconventional'' physical or computational
platforms such as analog electrical circuits \cite{he2019reservoir,hadaeghi2021neuromorphic}, optical media \cite{freiberger2017chip}, and chemical (molecular) substrates \cite{goudarzi2013dna}. It is, therefore, promising to design and implement functional sensors, processors, and controllers based on this computational framework. 

However, up to this point, the performance of RC in speech processing tasks has been relatively inferior to that of the deep-learning-based models \cite{triefenbach2013acoustic} and needs further improvements to fit conventional or end-to-end ASR systems with practical exploitation. In this regard, a major upgrade to shallow RC systems was introducing deep echo state networks (deep ESNs) that are able to capture temporal context of the input signal at different time-scales through several successively stacked RC layers \cite{ma2020deepr,gallicchio2017deep}. It enhanced the performance of RC in time-series prediction \cite{ma2020deepr}, short-term memory capacity (MC) task, and classification of experimental data in the field of computational biology \cite{gallicchio2017deep1}. However, it has remained to be further assessed if this alteration could also strengthen RC-based speech recognition models. In this study, therefore, we chose to explore applicability of RC models in speech recognition. Besides, since to the best of our knowledge, neither standard RC nor deep ESN have been employed to conduct any Persian ASR tasks, we trained conventional single-layer and deep ESNs to perform a speech recognition task on the Farsdat Persian speech dataset. 

\subsection{Related works}
\label{sec:Related works}

The FARSDAT dataset \cite{bijankhan1994farsdat} was collected in 1996 and since then has been used as the standard benchmark for developing Persian ASR systems such as Shenava \cite{Almasganj2001} and Nevisa \cite{sameti2008nevisa}. Similar to other ASR models, both systems comprise three modules for feature extraction, and acoustic and language modelling. Dimension reduction techniques such as principle component analysis, wavelet transforms, filter banks, Cepstral analysis, Mel frequency cepstrum, and kernel based methods are commonly employed to extract relevant features from each 20-30 ms frames (segments) of the speech signal. While standard Mel frequency Cepstral coefficient (MFCC) analysis was dominantly utilized in the feature extraction modules of the first editions of Persian ASRs, it has been recently suggested that LHCB (Logarithm of Hamming Critical Band filter banks) \cite{nejadgholi2009nonlinear}, convolutional neural networks (CNNs), and deep-belief-networks (DBNs) may return features that further increase the accuracy of Persian speech recognition in neural network models \cite{veisi2020persian}. 

In the acoustic and language modelling modules, combinations of Gaussian mixture models (GMMs) and deep neural networks with hidden Markov models have been extensively investigated to convert the extracted features to a probability over characters in the Persian alphabet, and to turn these probabilities into words \cite{firooz2017improvement,ansari2017toward,ansari2021rapid,asadolahzade2019improving}. In addition to conventional Persian ASR platforms, end-to-end systems based on modular deep neural architectures, and uni- and bi-directional long-short-term-memory (LSTM) framework were recently developed and tested on Farsdat dataset \cite{alisamir2018end,kermanshahi2021transfer,veisi2020persian}. The objective of current study is to investigate if reservoir computing models can be further integrated into either conventional or end-to-end Persian ASR systems. 

\subsection{Contributions}
\label{sec: Contributions}

We present, to the best of our knowledge, the first study that explores capabilities of reservoir computing in the context of Persian speech recognition. Specifically, we employed RC algorithms that are suitable for application to speech recognition. Moreover, we propose heterogeneous single and multi-layer ESNs to create non-linear transformations of the inputs that capture temporal context at multiple scales. The RC-based algorithms are also compared to a de-facto standard LSTM as problem-tailored state-of-the-art deep learning RNN solution.

\section{Model Architectures}
\label{sec:Model Architectures}
\subsection{Baseline architectures}
\subsubsection{LSTM}
 Long short-term memory RNN architectures are widely used for sequence labeling and prediction tasks, including speech processing applications such as spoken language translation \cite{sutskever2014sequence,cho2014learning} and speech recognition \cite{graves2013speech}. In LSTM networks, the standard hidden layer of recurrent neural networks (RNNs) has been replaced with  purpose-built gates and memory cells to filter and store the information. This modification has been proven particularly effective in tackling the vanishing gradient problem and fruitful in finding, memorizing, and exploiting long range dependencies in sequential data \cite{hochreiter1997long,graves2013speech}. In this study, we took the standard deep-learning LSTM architecture as de-facto standard in our speech recognition task.

\subsubsection{Shallow Reservoir Computing}
\label{ReservoirComputingModels}
\begin{figure*}[ht]
\centering
\includegraphics[scale=0.35]{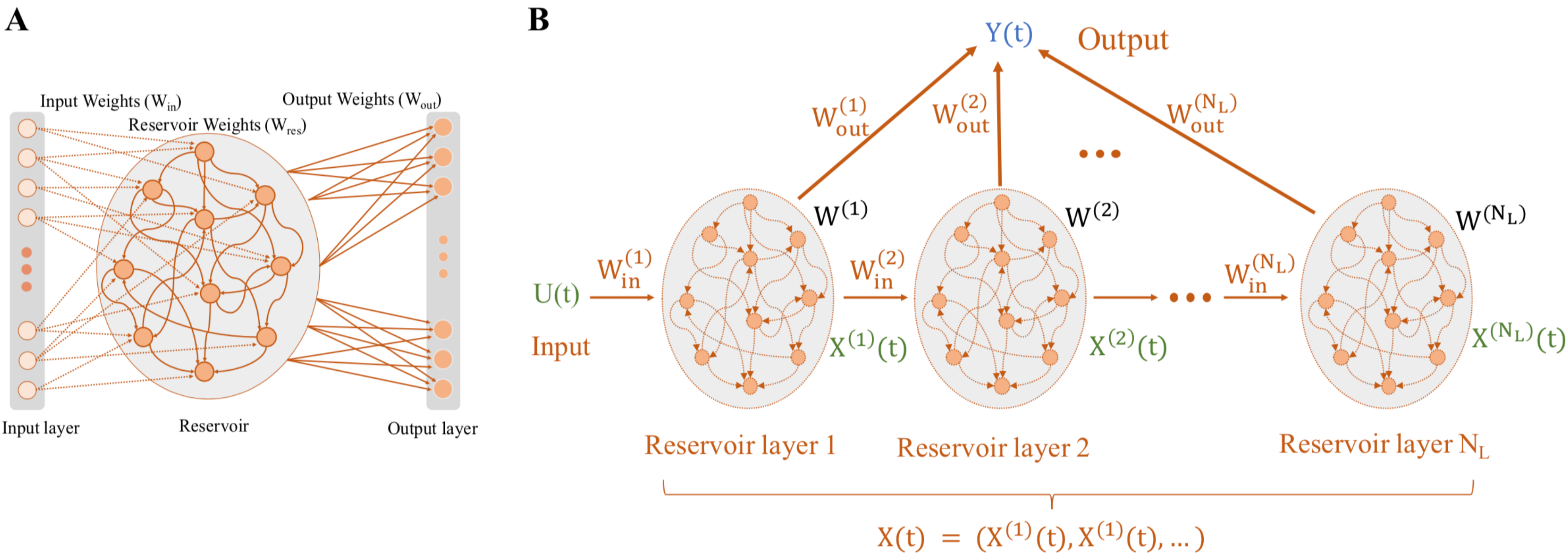}
\caption{Schematic illustration of \textbf{A:} a shallow and \textbf{B:} a deep reservoir computing model.} 
\label{fig: fig1}
\end{figure*}

Reservoir computing provides a computationally efficient framework for RNN design and training and has been successfully used in a large range of practical signal processing applications across different fields \cite{jaeger2004harnessing,lukovsevivcius2012reservoir,triefenbach2010phoneme}. 
% In speech processing tasks such as acoustic modeling, however, RC has proved promising by delivering comparable results to those offered by convolutional networks \cite{triefenbach2010phoneme}. 
The core to a typical reservoir computing model is a random, large, fixed recurrent neural network comprising a set of sparsely connected non-linear nodes (see Fig.~\ref{fig: fig1}.A). Through the internal
variables of this dynamical system (i.e., reservoir
states), the input signal presented to the RNN is non-linearly mapped into a higher dimensional signal space. These states are used to train a feed-forward readout module that is the only trained part of the network. The time-dependent output is computed as a linear combination of these random representations. Depending on the task, randomly
generated output to reservoir (all-to-all) feedback connections
may also be included in the architecture. 

Unlike traditional RNN training methods, the RC
technique proposes that the values of input-to-reservoir and reservoir connection weights are not critical and can be selected at random within some pre-defined intervals to obtain the best performance. Training only takes place in the readout layer where the signals from the individual nodes are fitted to a training signal, usually by a linear fit. Since only the output connections are trained and the optimization of the output layer only consists in a linear regression, reservoir computer can be faster and computationally more efficient than training a conventional recurrent neural network. 

The dynamics of a reservoir computer with real-time continuous value units is typically described by the following equations:
% \begin{equation}\label{eq: eq1}
% 	 \mathbf{x}(t+\Delta t) = f (\mathbf{W}\mathbf{x}(t)+\mathbf{W}^{in}\mathbf{u}(t+\Delta t)),
% \end{equation}

\begin{equation}
\label{eq: eq1}
\begin{split}
\mathbf{x}(t+\Delta t) & = (1- \eta a)\mathbf{x}(t) + \eta f (\mathbf{W}^{\text{in}}\mathbf{u}(t+\Delta t) + \mathbf{W}\mathbf{x}(t)+\mathbf{W}^{\text{fb}}\mathbf{y}(t)),
\end{split}
\end{equation}
where $\Delta t$ (here, $\Delta t = 1$) is the real time sampling period, $\mathbf{u}(t) \in \mathbb{R}^{N_{u}}$ is the input sequence, $\mathbf{x}(t) \in \mathbb{R}^{N_{x}}$ is the $N_{x}$-dimensional reservoir state, and $f$ is a nonlinear function. $\mathbf{W}\in\mathbb{R}^{N_x\times N_x}$, $\mathbf{W}^{in}\in\mathbb{R}^{N_x\times N_u}$, and  $\mathbf{W}^{\text{fb}} \in\mathbb{R}^{N_x\times N_y}$ are the input-to-reservoir, recurrent, and output-to-reservoir feedback weight matrices, respectively. In this equation, In this equation, $a>0$ and $\eta=1$ are reservoir neurons' leakage rate and time constant. The output, $\mathbf{y}(t)\in \mathbb{R}^{N_{y}}$ is then obtained from the extended system state (i.e., $ \mathbf{z}(t) = [\mathbf{x}(t); \mathbf{u}(t)] $ where $[.;.]$ stands for a vertical vector concatenation):
\begin{equation}\label{eq: eq2}
\mathbf{y}(t) = g(\mathbf{W}^{out}\mathbf{z}(t)),
\end{equation}
where $g$ is an output activation function and $\mathbf{W}^{out}$ is the readout weight matrix. Training the readouts is done by computing the linear regression weights of the target outputs on the harvested states of reservoir units and the inputs via pseudoinverse method: 

% With $N_{y}$ outputs, $N_{x}$ reservoir units and $N_{u}$ inputs, given the set $ \{y_{1i} , ..., y_{li}, z_{1i} , ..., z_{ki}\}_{k=N_{x}+N_{u}, l = N_{y}, i=1,...,N} $ of $N$ observations, the ordinary least square regression model is commonly used in standard RC to define the regression coefficients:
\begin{equation}\label{eq: eq3}
	 \mathbf{W}^{out} = \mathbf{Y}_{target}\mathbf{Z}^{T}(\mathbf{Z}\mathbf{Z}^{T} + \gamma^{2}I)^{-1}
\end{equation}
where $\mathbf{Z} $ is the matrix of extended system states, $\mathbf{Y}_{target}$ denotes the target sequence, $I$ is the identity matrix and $\gamma \geq 0$ is a regularization factor.

\subsubsection{Deep Reservoir Computing}
The study of hierarchically organized recurrent neural networks suggests that deep RNNs are able to progressively develop multi-scale representations of the temporal information in their internal states, making them attractive models for complex sequential learning tasks such as text, speech and language processing \cite{hermans2013training}. Such multi-layer hierarchical architecture comprises an input layer fed by the external stimuli, several intermediate recurrent network layers which receive the hidden state of the previous layer as input, and an output layer that only has access to the hidden state of the final hidden layer. The training process in these models entails computationally expensive techniques such as temporal back-propagation \cite{hermans2013training,gallicchio2017deep}, therefore, investigations in the filed of deep RNN are predominantly about the learning process. Deep reservoir computing, in contrast, provides a computationally efficient alternative approach for training. As depicted in Fig.~\ref{fig: fig1}.B, core components of this model are sequentially and unidirectionally stacked hidden layers where connections from higher to lower layers as well as connections from the input to layers other than the first level are avoided. Hidden states of these components are further exploited by a readout mechanism to learn the desired temporal task. Analogous to training a shallow ESN, in the case of a deep ESN, the state of the reservoirs at all the layers are concatenated in a predefined order to train a feed-forward readout module that is the only trained part of the network. 
Following the mathematical notation explained in \cite{gallicchio2017deep}, internal state update equation of the $i$-th layer of a deep ESN is as follows: 

\begin{equation}\label{eq: eq4}
\begin{split}
\mathbf{x}^{(i)}(t + \Delta t)& =(1-a^{(i)}) \mathbf{x}^{(i)}(t)+ a^{(i)} \tanh (\mathbf{W}_{i n}^{(i)} \mathbf{x}^{(i-1)}(t) +\boldsymbol{\theta}^{(i)}+{\mathbf{W}}^{(i)} \mathbf{x}^{(i)}(t)).
\end{split}
\end{equation}

Where $a^{(i)} \in [0,1]$ is the leaking rate parameter of the $i$-th layer. With a simple assumption that all the reservoir layers comprise $N_{R}$ neurons, $\mathbf{W}_{in}^{(i)}\in\mathbb{R}^{N_R\times N_R}$ and $\mathbf{W}^{(i)}\in\mathbb{R}^{N_R\times N_R}$ denote the input and recurrent weight matrices, respectively, and $\boldsymbol{\theta}^{(i)} \in \mathbb{R}^{N_{R}}$ is the bias-to-reservoir weight vector for this layer. Since the first layer (i.e., $i=1$) receives the external inputs (i.e., $\mathbf{x}^{(0)}(t) = \mathbf{u}(t)$), $\mathbf{W}_{i n}^{(1)} \in \mathbb{R}^{N_{R}} \times \mathbb{R}^{N_{u}}$ denotes a random input-to-reservoir weight matrix. 
Considering $l$ reservoir layers, one possible strategy to compute the output of a deep ESN at each time step, $t$, would be connecting the reservoir neurons in all the layers to the readout unit(s) and creating the extended system state as:

\begin{equation}\label{eq: eq5}
\mathbf{z}(t) = [\mathbf{u}(t); \mathbf{x}^{(1)}(t); \dots; \mathbf{x}^{(l)}(t)].
\end{equation}

Similar to the standard RC, the output at each time step, and the readout weights are further obtained by applying Eq.~\ref{eq: eq2} and Eq.~\ref{eq: eq3}. It should be noted that in this study, randomly
generated output-to-reservoir feedback connections were not included in the architecture. 

\subsection{Heterogeneous Reservoir Computing}
One technique to enhance the accuracy of the RC in ASR applications could be enforcing multi-scale spatiotemporal reservoir dynamics through increasing variability in the hidden layer. Therefore, different gradient-based learning methods have been previously employed to optimize global hyper-parameters as well as learnable neuron-specific variables to adjust timescales of this dynamical system such that the required memory to solve a given task is created \cite{jaeger2007optimization,manneschi2021exploiting,manneschi2021sparce,tanaka2021reservoir}. In this study, we propose a laminar-specific architecture to develop heterogeneous single and multi-layer ESNs which diversifies dynamics of individual units. 

In shallow ESN, we introduced unit variability by organizing the hidden layer into three sub-groups (i.e., laminar layers) of interconnected units  with slightly different state update equations. As denoted in Eqs.~\ref{eq: eq6} and \ref{eq: eq7}, the future state of reservoir nodes in each ensemble, $\mathbf{x}_{(i)}(t+\Delta t)$, is given in terms of the values of the other state variables at previous times:

\begin{equation}
\label{eq: eq6}
\begin{split}
\mathbf{x}(t+\Delta t) & = (1- a)\mathbf{\tilde{x}}(t) + f (\mathbf{W}^{\text{in}}\mathbf{u}(t+\Delta t) + \mathbf{W}\mathbf{\tilde{x}}(t)),
\end{split}
\end{equation}

where
% \begin{equation}\label{eq: eq6}
% 	 \mathbf{x}(t+\Delta t) = f (\mathbf{W}\mathbf{\tilde{x}}(t)+\mathbf{W}^{in}\mathbf{u}(t+\Delta t)),
% \end{equation}

\begin{equation}\label{eq: eq7}
\mathbf{\tilde{x}}(t) = [\mathbf{u}(t); \mathbf{x}_{(1)}(t-\tau_{1}); \dots; \mathbf{x}_{(l)}(t-\tau_{l})].
\end{equation}

In this notation, $\mathbf{x}_{(i)}$ stands for $i$-th pre-defined sub-group of internal neurons. It should be noted that the recurrent weight, $\mathbf{W}\in\mathbb{R}^{N_x\times N_x}$, and the input-to-reservoir weights, $\mathbf{W}^{\text{in}} \in \mathbb{R}^{N_x\times N_u} $, are initiated randomly before assigning multiple time delays to different sub-groups. 
% In our case, there exists three sub-groups with $\tau_{i} \in {1, 3, 5}$, where, for instance, the current state of the nodes in the first ensemble depends on the state variables of the same ensemble at previous times and the state variables of other two ensembles at $t-3$ and $t-5$, respectively.

For deep ESN architecture, we applied the same multiple timescale framework by assigning gradually increasing time delays to successive layers in a way that the state variables in deeper layers depend on a longer history of their own states. To implement that, we simply modified the state equation and the extended system introduced in Eqs.~\ref{eq: eq4} and \ref{eq: eq5} as follows:

\begin{equation}\label{eq: eq8}
\begin{split}
\mathbf{x}^{(i)}(t + \Delta t)& =(1-a^{(i)}) \mathbf{x}^{(i)}(t-\tau_{i}) +
a^{(i)} \tanh (\mathbf{W}_{i n}^{(i)} \mathbf{x}^{(i-1)}(t) +\boldsymbol{\theta}^{(i)}+{\mathbf{W}}^{(i)} \mathbf{x}^{(i)}(t-\tau_{i})).
\end{split}
\end{equation}

\begin{equation}\label{eq: eq9}
\mathbf{z}(t) = [\mathbf{u}(t); \mathbf{x}^{(1)}(t-\tau_{1}); \dots; \mathbf{x}^{(l)}(t-\tau_{l})],
\end{equation}

where $\tau_{i}$, the internal delay in $i-$th ensemble/layer, can be either fine-tuned as a hyperparameter or trained via gradient based methods. 
\section{Materials and methods}
\subsection{Data-set and data characteristics}

All the experiments in this manuscript were conducted on the FARsi Speech DATabase (FARSDAT)\cite{bijankhan1994farsdat} which has been created for Persian speech and speaker recognition purposes. This collection comprises 608 wave files, consisting $20$ seconds long of $9-12$ sentences spoken by 304 native Persian speakers with non-identical accents from different age groups. FARSDAT audio signals were recorded with sampling rate of 22.5 KHz, and the signal-to-noise ratio is 34 dB. Manual annotations at phoneme as well as word-level have been provided for all utterances. 

Among the uttered sentences, two are common to all speakers and contain all the Persian alphabet letters except for ``\textit{fe}''. In the experiments conducted in this article, recordings of these two sentences uttered by 100 speakers were adopted for reporting the primary results of each RC structure, as well as, hyper-parameter optimization. Therefore, within one-fold cross-validation setting, the training set includes the speech signal of 70 speakers (20 percent of which was randomly selected for validation) and the test set contains audio signals recorded from another 30 speakers. 
In addition to the performance on the small set of common sentences, we report the recognition results on total FARSDAT speech database, which we refer to as \textit{complete FARSDAT} database where the total sentences uttered by 297 speakers (i.e. 5940 sentences) were selected as the training set and those recorded from the other speakers (140 sentences) were utilized to test the models.

\subsection{Pre-processing}
During the experiments, each audio signal was segmented to overlapped frames of the same lengths (usually 23 ms with 12.5 ms overlap) to ensure the signal segments remain statistically static. The frames were further windowed by the Hamming window to calculate the spectrum of the frame using the Fast Fourier Transform (FFT). The received spectrum was afterward passed through the 18 filter banks on the Bark scale. Then, the logarithms of the energies under each filter were separately calculated to obtain 18 LHCB (Logarithm of Hamming Critical Band filter banks) representation vectors \cite{nejadgholi2009nonlinear}. Since, LHCB features are sensitive to noise and signal variations, the longitudinal norm-2 (i.e., mean and variance normalization) was used to normalize LHCB components to enhance the accuracy and efficiency of the developed models. 

The features extracted from training data were afterward exploited to adjust models' hyper-parameters and to train learnable parameters of the neural networks. Finally, the obtained models were evaluated on two test data, the small set of common sentences and the \textit{complete FARSDAT}.

\subsection{Evaluation metrics}
In this study, the accuracy of the models on test sets is measured at the frame level. Frame recognition rate is computed based on the frame-wise comparison of the reference labels for each frame with model predictions. 
% Phone recognition rate counts correctly predicted phonemes corresponding to the input signal. This value is calculated as sum of the number of insertions, deletions, and substitutions needed to convert the recognized (predicted) phone string correspond to the input signal to the phone target (reference) sequence. 

\section{Experiments and results}

All our RC models are grounded on a reservoir of leaky-integrator neurons \cite{jaeger2007optimization} with a few control parameters whose values determine the emergent behavior of the system in response to the features extracted from the input audio signals. Number of the input units is determined by the number of extracted features and the extent of consecutive frames that provide the phoneme recognition networks with the acoustic context of each input sequence. In this study, we extracted 18 features from the input frames and considered 14 consecutive frames for each sentence to present the RC models with $18*14$ input sequences at a time. That is, the recurrent neural network slides on the sequence of features extracted from each sentence and updates its states. At each time point, a sequence of 14 components from feature vectors are fed to the network that has to recognize the phoneme label of the central vector. All models comprise 35 output units representing 35 Persian phonemes. 

We implemented our models using the commercial toolbox MATLAB 2017b (computer hardware: Intel Core i7 (2.7 GHz), 16 GB RAM, NVIDIA GeForce GT 650M).

\subsection{Shallow RC}
\begin{figure}[ht]
\centering
\includegraphics[scale=0.65]{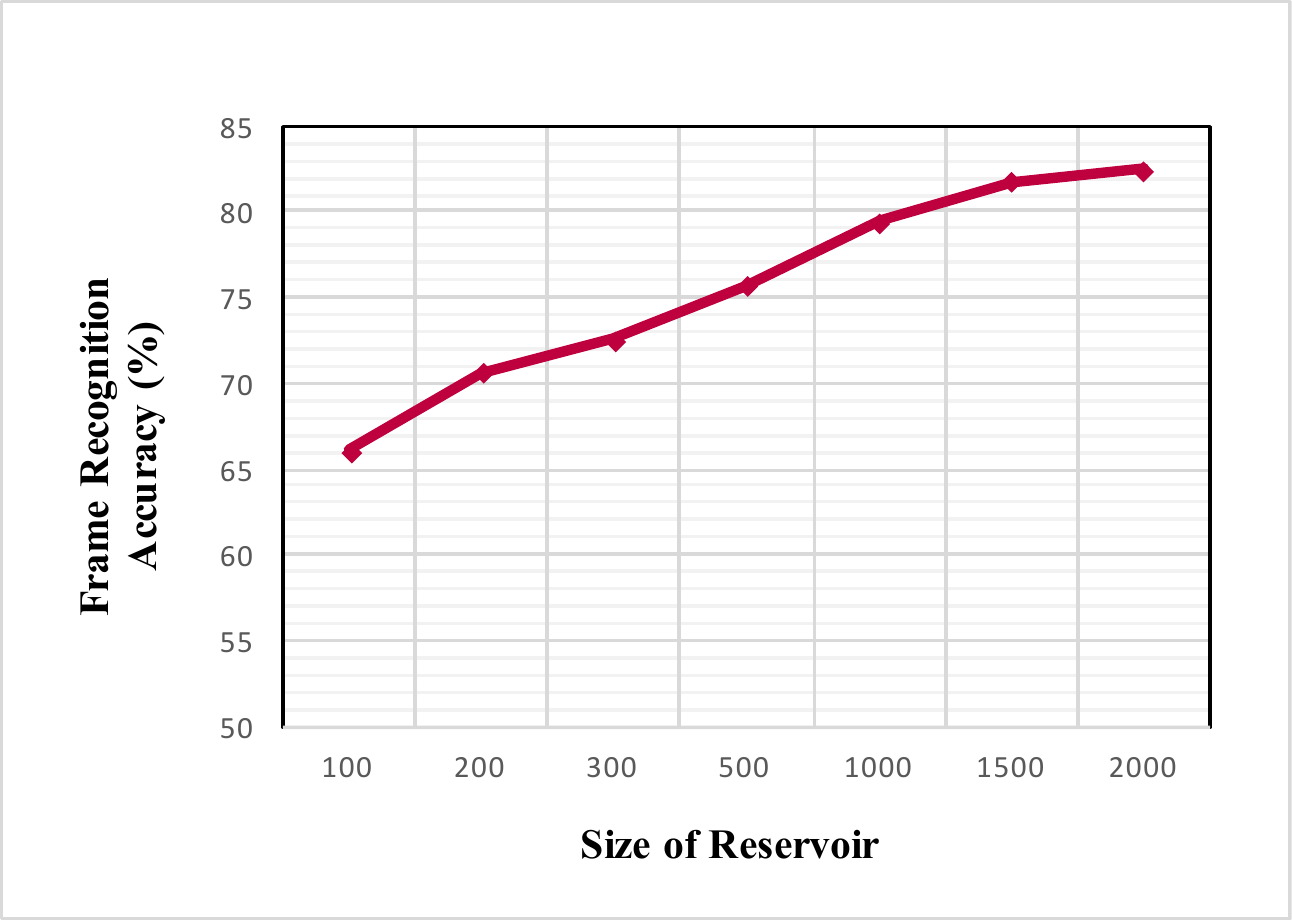}
\caption{Effect of the reservoir size in the performance of the shallow RC as the speech recognition model. The models were trained on speech signal of 56 speakers and validated on audio signals recorded from 14 speakers.} 
\label{fig: fig2}
\end{figure}

In our experiments on shallow RC, we gradually increased the size of the reservoir to examine the role of this parameter in the performance of the network. The activation function of the reservoir units and the output layer were considered as tangent hyperbolic and identity functions, respectively. We calculated the readout weights through the pseudo inverse method. Both input-to-reservoir and reservoir internal weights were sampled from a uniform distribution. Extensive grid searches were conducted to optimize hyper-parameters of the model. As a result, we set spectral radius, $\rho = 0.3$ and the leakage rate, $a = 0.5$. We also sampled input-to-reservoir weights from a uniform distribution with real values between $[-0.1,~ 0.1]$. Since the initial values for the input-to-reservoir and reservoir internal weights are random, for each reservoir hyper-parameter, the training and testing of the RC were repeated 5 times and the average results were reported (see Fig.~\ref{fig: fig2}).

As expected, increasing the size of the hidden layer enhances the performance such that the largest network with 2000 neurons achieved 82.52\% frame recognition accuracy. 

\subsection{Deep RC}
\begin{figure}[ht]
\centering
\includegraphics[scale=0.58]{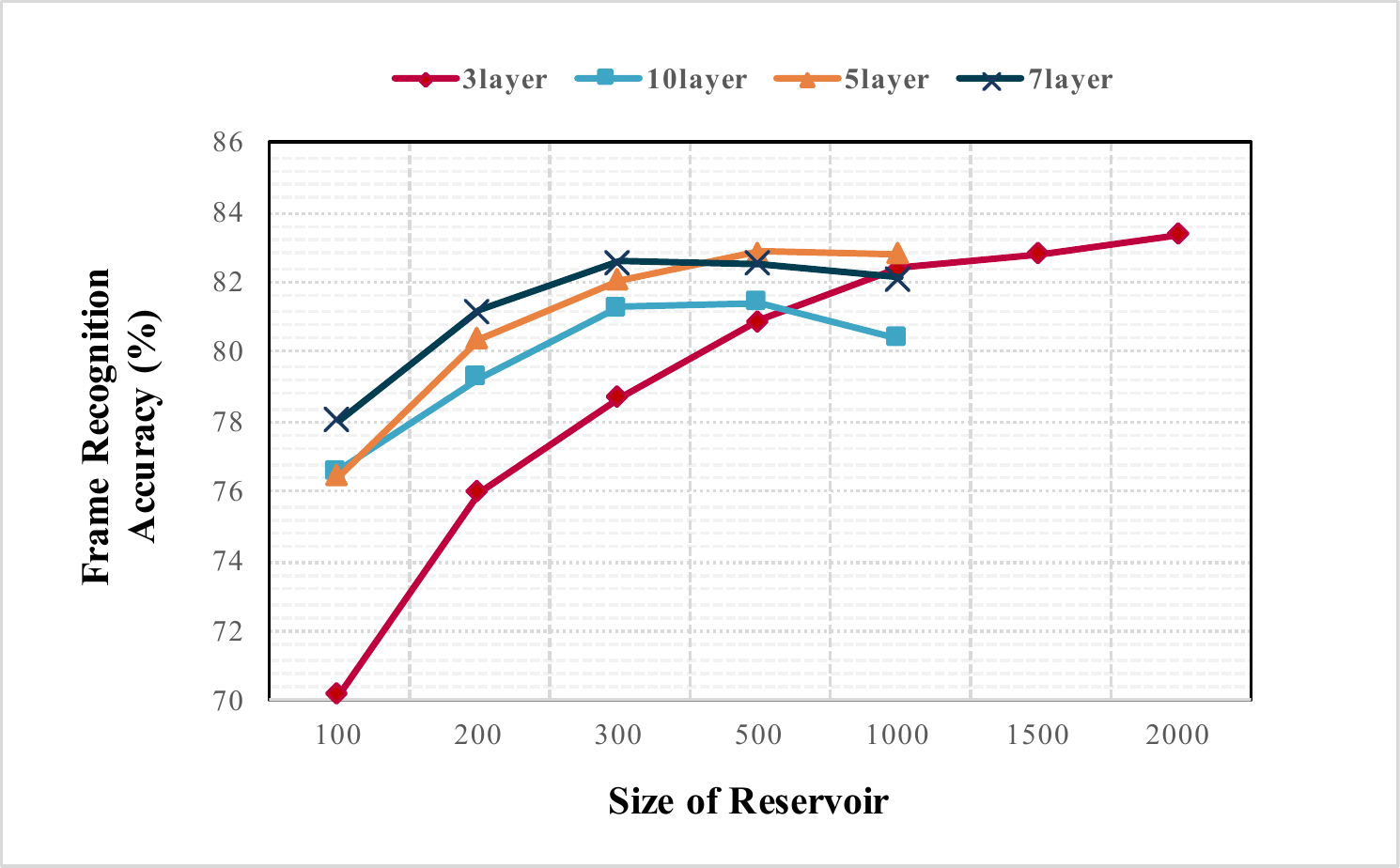}
\caption{Effect of the reservoir size in the performance of the deep ESN as the speech recognition model. The models were trained on speech signal of 56 speakers and validated on audio signals recorded from 14 speakers.} 
\label{fig: fig3}
\end{figure}

Afterward, we investigated the effect of stacking hidden layers in frame recognition accuracy. To this end, different deep RC structures with $3, 5, 7$ and $10$ hidden layers were implemented. For the sake of simplicity, we used the same reservoir size and the hyper parameter values for all hidden layers. Again through extensive grid search, we optimized spectral radius and the leakage rate for each structure. Fig.~\ref{fig: fig3} depicts frame recognition rates of different deep RC structures with varying reservoir sizes. Noticeably, the best accuracy, 83.38\%, is obtained from a 3-layer deep ESN with 2000 neurons in each layer, showing the potential overfitting effect after stacking more layers. In the case of $5, 7$ and $10$ hidden layers, after the reservoir size exceeds 1000 neurons, the recognition rate declines. Therefore, we didn't report the frame recognition rate of the larger reservoir sizes. 

% the best accuracy, 83.38, is attained by the 3 hidden-layer RC with the reservoir size of 2000 for each layer and the spectral radius 0.3, leakage range 0.5 and the input weight matrix in the range of [-0.1-0.1].

\subsection{Heterogeneous RC}
Our experiments on standard shallow and deep RC models suggest that organizing the model in a hierarchical structure with three layers only enhances the frame recognition rate by $0.27\%$. Therefore, we, firstly, sought to see if introducing heterogeneity to shallow architecture can show comparable impact. Then, we investigated the effect of layer variability in deep RC structures. As reported in Table.~\ref{tbl: tbl1} presence of multiple time-scales prolongs the internal memory and further improves the performance of the network in speech recognition. In fact, with only 2000 neurons, it outperforms the performance of the deep ESN with 6000 internal units.

In this study, we only considered three sub-groups with $\tau_{i} \in \{1, 3, 5\}$ and studied the effect of number of nodes in each population on the performance. Our experiments showed that, the best result (reported in Table.~\ref{tbl: tbl1}) is obtained when all ensembles had identical sizes. The underlying rationale for the choice of $\tau_{i} \in \{1, 3, 5\}$ was that on the one hand, introducing longer time lags enables RNNs to incorporate extra temporal context which leads to better frame recognition scores. On the other hand, however, RNNs seem to only benefit from having access to a limited range of previous contextual information in framewise phoneme classification (e.g., see Fig.3 in \cite{graves2005framewise}). Therefore, we pre-set the lags and optimized the size of each population. 

\begin{figure}[ht]
\centering
\includegraphics[scale=0.6]{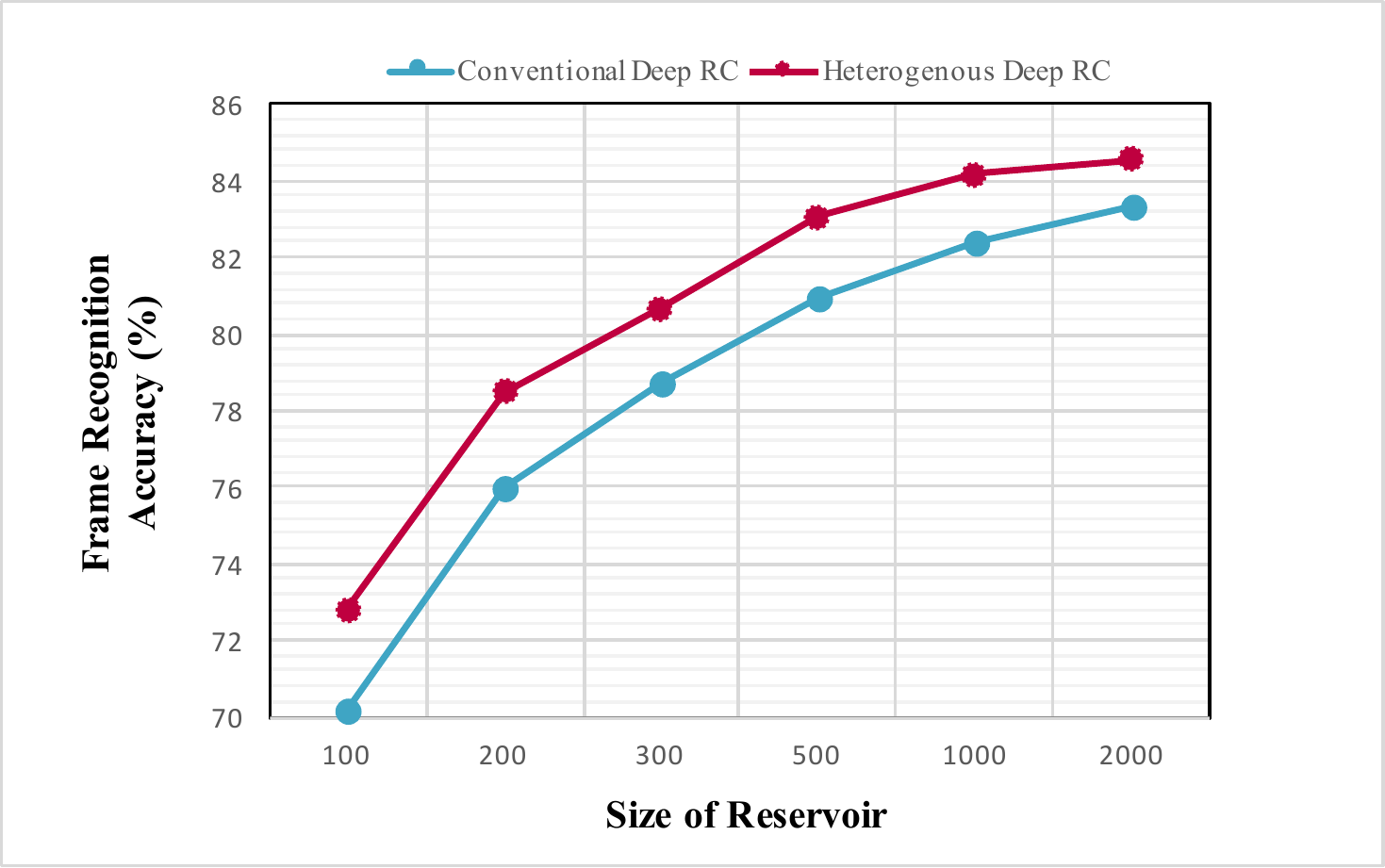}
\caption{Effect of the proposed heterogeneity in the performance of deep ESN models. The models were trained on speech signal of 56 speakers and validated on audio signals recorded from 14 speakers.} 
\label{fig: fig4}
\end{figure}

We also explored heterogeneous deep ESN networks where intermediate layers have non-identical temporal scales. We assigned $\tau_{i} = \{1, 3, 5\}$ to superficial, intermediate and deep layers, respectively. Again, we studied networks with various identical and non-identical layers and noticed the best performance is obtained with a 3 layer network where each layer comprises 2000 internal nodes. As shown in Fig.~\ref{fig: fig4} and reported in Table.~\ref{tbl: tbl1}, the proposed heterogeneity improves recognition accuracy of the deep ESN by approximately $1.25\%$ on small dataset.   

Conducting preliminary experiments on the \textit{complete FARSDAT} database, consistency of these findings were confirmed (see Table.~\ref{tbl: tbl1}). However, in this study, we only tested RC models with 500 neurons.

\begin{table}[htb]
\centering
\caption{Performance evaluation of the reservoir computing models in terms of frame recognition accuracy. Test sets are both small and complete FARSDAT datasets.}
\label{tbl: tbl1}
\vspace{5pt}
\resizebox{0.6\columnwidth}{!}{%
\begin{tabular}{l c c}
\toprule
    & \small{\textbf{small FARSDAT}}        & \small{\textbf{complete FARSDAT}}  \\
\toprule
\small{\textbf{Shallow ESN}}  & 82.05\% & 67.66\%  \\
\small{\textbf{Deep ESN}}  & 82.32\% & 71.86\%  \\
\small{\textbf{Heterogeneous shallow ESN}} &82.72\% & 68.74\%  \\
\small{\textbf{Heterogeneous deep ESN}}  & 83.57\% & 72.51\% \\
% \textbf{LSTM}  & 84.99\% & 84.26\%  \\
\hline
\end{tabular}
}
\vspace{-10pt}
\end{table}

\subsection{LSTM}
We compared RC speech recognition performance with an LSTM network with three hidden layer of $100$ LSTM cells and a Softmax readout layer. Similar to RC models, we presented the models with sequences of feature vectors extracted from consecutive frames representing spoken sentences. This network was trained using the stochastic gradient descent (SGD) with Momentum method and the learning rate of $0.0001$. On the small and complete FARSDAR dataset, the mean frame recognition rates are 84.99\% and 84.26\%, respectively.

\section{Discussion and conclusions}
The present study explores the potentials of reservoir computing (RC) for Persian speech recognition as a common practice in ASR applications. The underlying rationale was that RC represents a computationally efficient RNN-based approach to learn temporal features. Our results suggest that the RC models perform in terms of frame recognition accuracy at least on par with the de-facto standard in speech processing, the LSTM. It should, however, be noted that the implemented RC models have a vastly fewer trainable parameters (approximately $70000$ in the current study) than the LSTM ($>305500$). In the current experiment, this led to a reduction of training time from 150 minutes for the LSTM to 90 minutes for the RC system. It should also be mentioned that LSTM training was performed on GPU and was already highly optimized for GPU usage, while the RC training was on CPU and was not optimized for parallel computing.

Moreover, introducing heterogeneity in terms of variable time-scales further improved recognition accuracy of both shallow and deep ESN up to $1.25\%$. On the complete FARSDAT dataset, however, we preliminary tested our RC models with only 500 neurons. Therefore, the performance is not comparable to LSTM. It remains to be shown that our models can be improved by further increasing the size of hidden layers. As a further extension to this study, the effect of additional number of heterogeneous sub-groups and different values of temporal delays in both shallow and deep ESN will be systematically explored. Besides, in this study, we only reported the performance in terms of frame recognition accuracy. It remains for future works to conduct recognition tasks at phoneme and word levels.  

\section{Acknowledgments}

Fatemeh Hadeaghi's research was supported by the Deutsche Forschungsgemeinschaft, Germany (TRR 169/A2).

\bibliographystyle{IEEEtran}
\bibliography{References.bib}

\end{document}